\documentclass{amsart}
\usepackage[dvipdfmx]{graphicx}
\usepackage{amsmath,amssymb,mathrsfs,hyperref,subcaption}
\newcommand{\bvec}[1]{\mbox{\boldmath $#1$}}
\newcommand{\vect}{\mathbf}
\newcommand{\REVIEW}[4]{\textit{#1} \textbf{#2},#4 (#3).}
\newcommand{\Name}[1]{#1,}

\title{Unidirectional control of optically induced spin waves}
\author{I. Yoshimine${}^{1,2,4}$}
\author{Y. Y. Tanaka${}^1$}
\author{T. Shimura${}^1$}
\author{T. Satoh${}^{1,3}$}

\thanks{${}^1$Institute of Industrial Science, The University of Tokyo - Tokyo 153-8505, Japan}
\thanks{${}^2$RIKEN - Sendai 980-0845, Japan}
\thanks{${}^3$Department of Physics, Kyushu University - Fukuoka 819-0395, Japan}
\thanks{${}^4$E-mail: isao.yoshimine@riken.jp}

\begin{document}
\begin{abstract}
	Unidirectional control of optically induced spin waves in a rare-earth iron garnet crystal is demonstrated. We observed the interference of two spin-wave packets with different initial phases generated by circularly polarized light pulses. This interference results in unidirectional propagation if the spin-wave sources are spaced apart at $1/4$ of the wavelength of the spin waves and the initial phase difference is set to $\pi /2$. The propagating direction of the spin wave is switched by the polarization helicity of the light pulses. Moreover, in a numerical simulation, applying more than two spin-wave sources with a suitable polarization and spot shape, arbitrary manipulation of the spin wave by the phased array method was replicated.
\end{abstract}

\maketitle
Spin waves---collective modes of spin precessions---are considered promising information carriers in the field of magnonics, as Joule heating is negligible and propagation damping is low\cite{Kruglyak2010,Chumak2015,Gurevich1996,Stancil2009,Kajiwara2010}. For various technologies such as spin-wave switching\cite{Lenk2011}, spin-wave-assisted recording\cite{Lenk2011,Seki2013}, and sensing of small magnetic fields\cite{Lee2010}, spatial control of the spin wave is indispensable. This was demonstrated in previous studies by employing microwaves\cite{Buttner2000,Covington2002,Perzlmaier2008,Tamaru2011} or waveguides\cite{Kanazawa2016,Wang2002,Vogt2014,Haldar2016,Wagner2016}. Nevertheless, for these methods, adjusting the electrodes or waveguides is required to modify spin-wave direction or position. Recently, optically induced spin waves were reported for which propagation direction and wavelength of the spin wave were varied by spatially shaping the pump spots\cite{Satoh2012}. Furthermore, the initial phase of the spin wave can be modified via the polarization of the light pulse\cite{Satoh2012,Yoshimine2014}. As there is no need to attach electrodes or antennas to media, one can readily change the position of the spin wave excited by the light pulses.

In our study, the unidirectional propagation of the spin wave was realized through interference of two spin waves with different initial phases. Additionally, we numerically demonstrated various spatial manipulation of spin waves by applying a phased array method that has been employed in microwave radar sensing\cite{McManamon1996,Yao2009}, flaw detection using ultrasonic wave\cite{McNab1987}, and manipulation of surface plasmons\cite{Kosako2010,Tanaka2015,Lin2013} and phonons\cite{Feurer2003}.

Generally, simultaneous generation of two plane waves with a $\pi /2$ phase difference and a spatial gap of $1/4$ wavelength leads to unidirectional propagation. When the waves are generated at $x=+d$ and $x=-d$ with the initial phase $+\delta$ and $-\delta$, respectively, the one-dimensional waveforms $E$ at $x>+d$ and $x<-d$ are written
\begin{align}
E(x>d) &= A \exp \left[ i(\omega t-k(x-d)+\delta) \right]+ \notag\\
&\qquad A \exp \left[ i(\omega t-k(x+d)-\delta) \right] \notag\\
&=A \exp \left[ i(\omega t-kx)\right] \times (\exp \left[i(kd+\delta) \right] \notag\\
&\qquad +\exp \left[ i(-kd-\delta) \right] ), \label{rightside}\\
E(x<-d) &=A \exp \left[ i(\omega t-k(d-x)+\delta) \right]+ \notag\\
&\qquad A \exp \left[ i(\omega t-k(-d-x)-\delta) \right] \notag\\
&=A \exp \left[ i(\omega t+kx) \right] \times (\exp \left[ i(-kd+\delta) \right] \notag \\
&\qquad +\exp \left[ i(kd-\delta) \right] ), \label{leftside}
\end{align}
where $k$ is wavenumber, $\omega$ is angular frequency, $A$ is amplitude.
If $2d=\lambda/4=\pi/2k, 2\delta = \pi/2$, it ends up with
\begin{align}
   E(x>d) &=0, \\
   E(x<-d) &=2A\exp \left[ i(\omega t+kx) \right].
\end{align}
As a result, the amplitude is doubled at $x<-d$ and zero at $x>+d$. Hence, two spin-wave sources separated by 1/4 of the wavelength with an initial phase difference of $\pi/2$ can produce unidirectionally propagating spin waves.

In magnetically ordered materials, spins excited by the light pulses start precessing, thereby forming a spin wave. For circularly polarized light pulses, the interaction between spins and light pulses is known as the inverse Faraday effect\cite{Kimel2005}. The initial phase of the precession can be shifted by $\pi$ through the polarization helicity.

\begin{figure}
   \begin{center}
      \includegraphics[width=80mm]{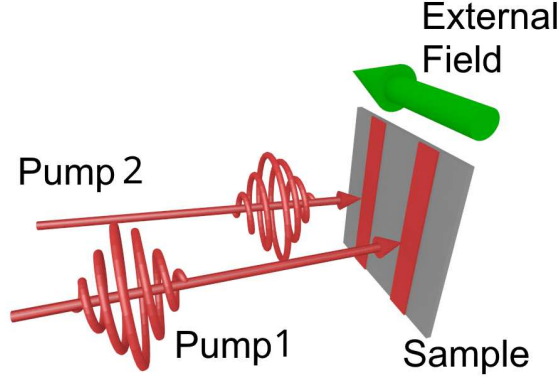}
   \end{center}
   \caption{Experimental setup: Circularly polarized pump pulses are focused onto two parallel stripes on the sample.}
   \label{fig:setup}
\end{figure}

In a bismuth-doped rare-earth iron garnet crystal $(\textrm{Gd}_{1.5}\textrm{Yb}_{0.5}\textrm{Bi})\textrm{Fe}_{5}\textrm{O}_{12}$ of thickness 120~$\mu$m, we generated two spin-wave packets and observed the propagation of the resulting spin wave by the pump-probe method. In-plane external field of 1000 Oe was applied to make the sample monodomain. As shown in Fig.~\ref{fig:setup}, a pair of pump pulses (pump1 and pump2) were focused on a pair of left and right parallel stripes with width of 110~$\mu$m and length of 4.6~mm. The width and length of the light stripes were determined from $d$ in the equation, $I(r)=\exp (-2r^2/d^2)$; here $I$ is intensity. The pump pulse durations were 150~fs, and the center wavelengths were 1300~nm. 
The typical center wavelength of the spin waves generated by the pump pulses was 250~$\mu$m. Therefore, for unidirectional control of propagation, the spacing between the two stripes was set to 60~$\mu$m, which is roughly $1/4$ of the center wavelength.

\begin{figure}[tb]
   \centering
   \includegraphics[width=75mm]{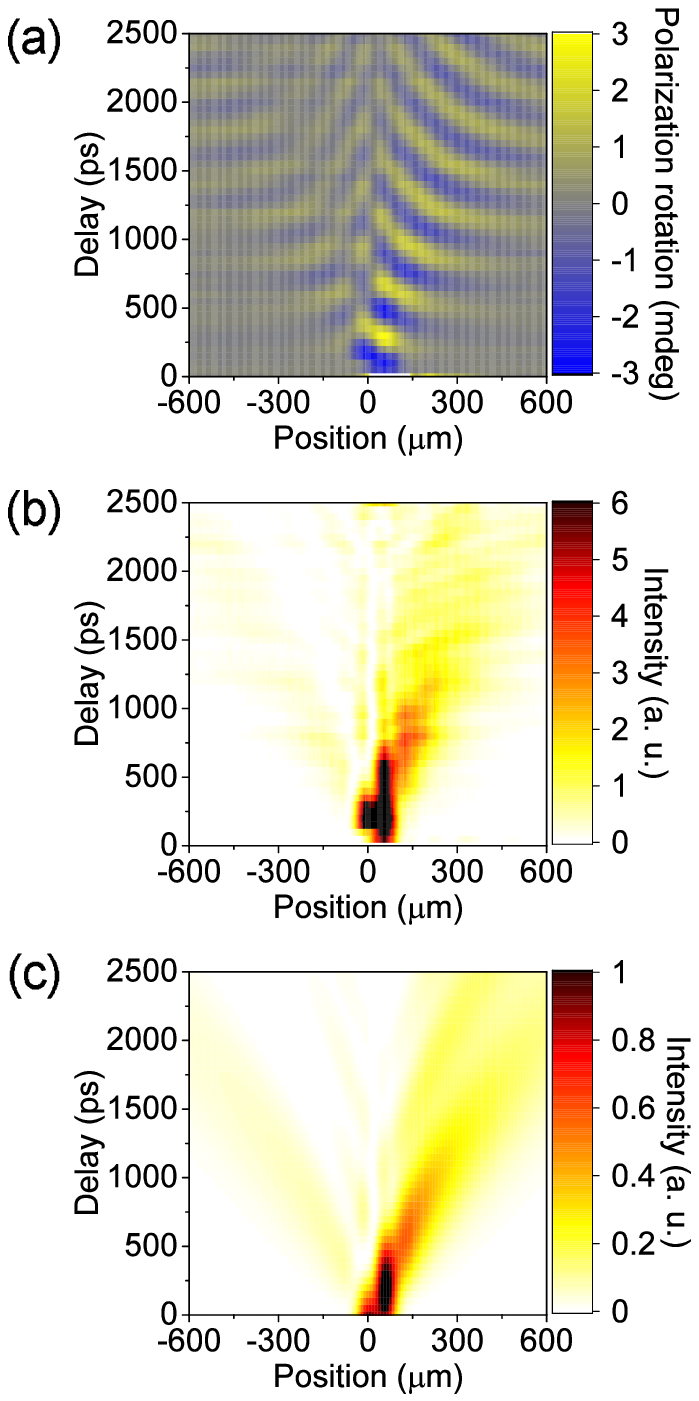}
   \caption{Experimental and numerical results of spin-wave interference. The spatiotemporal plots present spatial position along the abscissa and time delay for spin-wave generation from the right stripe along the ordinate: (a) waveform of the propagated spin wave, (b) spin-wave intensity, and (c) calculated spin-wave intensity. The origin is set at the center of pump1.}\label{fig:wfright}
\end{figure}

A $\pi/2$ phase difference was achieved by applying a temporal delay in the pulse of pump1 with respect to the pulse of pump2. The center frequency of the spin waves was 2.5~GHz; hence the temporal delay was set to 100~ps. We obtained a spatiotemporal image of the spin wave from the Faraday rotation of the probe polarization with a combination of a CCD camera and a polarizer\cite{Yoshimine2014,Parchenko2013}. The Faraday rotation was measured from 21 pictures taken with different orientation of the axis of the polarizer positioned in front of the camera. For each picture, the orientation of the axis was different by 0.1~degrees. The resulting spatiotemporal waveform and intensity of the spin wave are presented in Fig.~\ref{fig:wfright}(a) and (b), respectively.

For comparison, we calculated the spatiotemporal waveform of spin wave $m(\bvec{r},t)$ as
\begin{align}
   m(\vect{r},t)=\int_k h(\vect{k})\exp [ i(\omega(\vect{k})t &-{k}\cdot \vect{r}+\xi) ] \notag \\
   &\times \exp \left[ -\alpha \omega (\vect{k})t \right] d \vect{k}. 
\end{align}
Here $h(\vect{k})$ is the Fourier transform of the pump spot shape $h(\vect{r}) $, which is assumed Gaussian,
\begin{align}
h(\vect{r})=\exp \left[ \frac{-r^2}{2r_0^2}\right]
\end{align}
and $ r_0 = 35~\mu$m.
The sum of the calculated $m(\vect{r},t)$ for spin waves generated from two stripes results in a spatiotemporal waveform corresponding to interfering spin waves. The dispersion relation $\omega (\vect{k})$ of the sample was calculated with a thickness of 120~$\mu$m, an out-of-plane anisotropy field of 600~Oe, saturation magnetization 4$\pi M_\mathrm{s}$ of 1210~G, and an external field of 1000~Oe. The calculated spin-wave intensity map [Fig.~\ref{fig:wfright}(c)] exhibits a strong peak at the right of the pump spots, and hence the spin waves are propagating to the right. This result is in good agreement with the experimental result [Fig.~\ref{fig:wfright}(b)]. Thus, unidirectional propagation of spin waves was realized with light pulses.

\begin{figure}[tb]
   \centering
   \includegraphics[width=75mm]{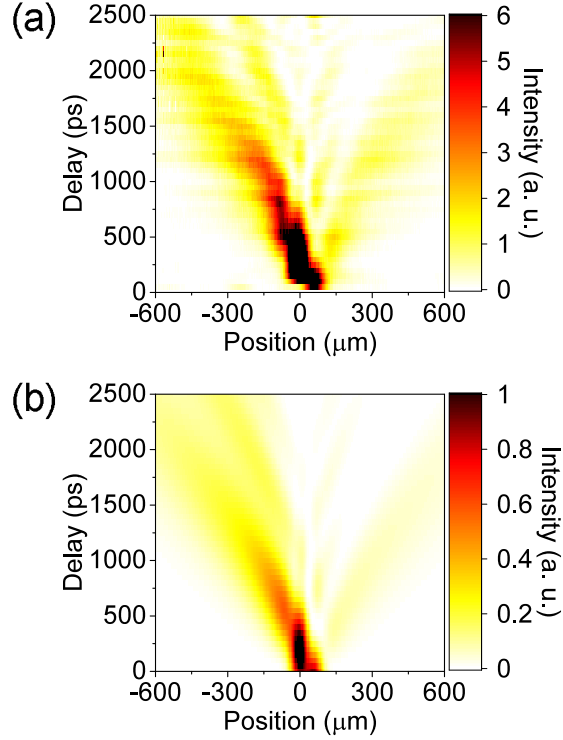}
   \caption{Spatiotemporal plot of the spin-wave intensity when polarization helicity is reversed: (a) experimental and (b) calculated results.}\label{fig:wfleft}
\end{figure}

If $2 \delta = -\pi/2$ instead of $\pi/2$ in Eqs.~(\ref{rightside}) and (\ref{leftside}), then
\begin{align}
E(x>d) &=2A\exp \left[i(\omega t-kx)\right], \\
E(x<-d) &=0,
\end{align}
meaning that if the sign of the mutual phase between the two spin waves is reversed, the propagation direction should be also reversed. This prediction was examined experimentally and by numerical simulation. 
To reverse the sign of the mutual phase between the two spin waves in the experiment, the polarization helicity of pump2 is reversed. 
The experimental and numerical results of the spin-wave intensity [Figs.~\ref{fig:wfleft}(a) and (b), respectively] are in good agreement. Note that the spin waves propagate to the left. Thus, we have shown that the unidirectional propagation direction of the spin wave is switchable by taking advantage of the polarization helicity of the pump pulses.

\begin{figure}[tb]
   \begin{center}
      \includegraphics[width=65mm]{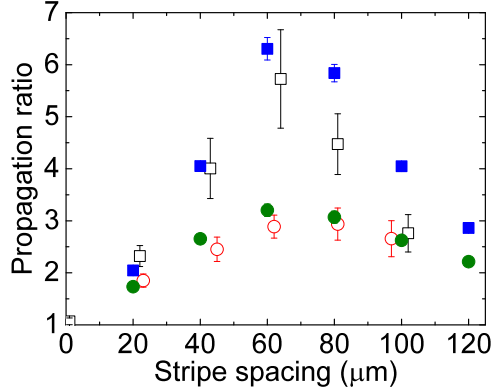}
   \end{center}
   \caption{Stripe spacing dependence of the propagation ratio. Squares (black open: experiment, blue filled: simulation) show propagation to the right. Circles (red open: experiment, green filled: simulation) show propagation to the left.}
   \label{fig:distance}
\end{figure}

Next, we investigated the dependence on stripe spacing of the propagation ratio, which is defined as the ratio of spatiotemporally integrated intensity of the spin waves propagating to the right ($0$--$600~\mu$m) to that propagating to the left ($-600$--$0~\mu$m) in Fig.~\ref{fig:wfright}. The definition of the ratio is reversed when the polarization helicity of the pump pulses is reversed, as shown in Fig.~\ref{fig:wfleft}.  
The temporal integration range was 1.0--1.4~ns. We measured and calculated the propagation ratio for the stripe spacing of 40, 60, 80, 100, 120 $\mu$m. From the experimental and numerical results (Fig.~\ref{fig:distance}), the highest propagation ratio was obtained at 60~$\mu$m. There is a distinct difference in the propagation ratio between spin waves propagating to the right and those propagating to the left. This difference stems from the temporal delay in spin-wave generation. In the experiments, spin waves generated from the right stripe (pump2) were generated 100~ps earlier than those generated from the left stripe (pump1). Therefore, to the right of the stripe pair, the difference in amplitude of the two spin waves is larger than that to the left of the stripe pair. This means that if the spin waves interfere destructively, a spin wave of some amplitude remains. This amplitude becomes larger if destructive interference occurs on the right side rather than the left. This leads to a decline in the propagation ratio of the spin wave that propagates to the left.

\begin{figure}[tb]
   \centering
   \includegraphics[width=80mm]{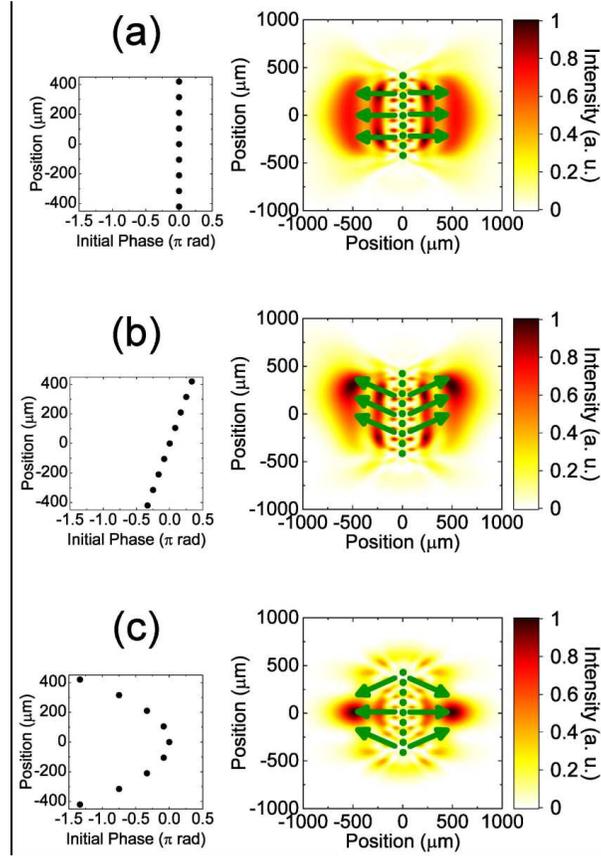}
   \caption{Numerical results of spatial spin-wave manipulation by a phased array (9-spot array). The graph (left) gives the initial phase distribution for the intensity plot (right) at 2~ns after the spin-wave generation. Green spots indicate positions of pump spots, and green arrows show the direction of the spin-wave propagation. (a) Same phase: propagation is perpendicular to the array, (b) Linear phase shift: propagation is tilted, (c) Parabolic phase shift: spin wave converges.}\label{fig:phasedarray}
\end{figure}

The above-mentioned technique can be further extended to spatial manipulation of spin waves by the phased array method. Here we demonstrate this in numerical simulations. In the simulation, the propagation of nine spin waves generated in different position and initial phase was calculated separately and superposed to obtain the spatial waveform. The results for the interference of nine spin-wave packets having the same initial phase [Fig.~\ref{fig:phasedarray}(a)] show a spin wave propagating in the direction perpendicular to the pump spot array. In contrast, if a linear phase difference is applied, the propagation direction is tilted from the perpendicular direction [Fig.~\ref{fig:phasedarray}(b)]. Moreover, if a parabolic phase shift is applied, the spin wave converges [Fig.~\ref{fig:phasedarray}(c)].

From the anisotropic dispersion of the spin waves in rare-earth iron garnet, the initial phase distribution and the resultant waveform has a non-intuitive relationship. For the parabolic phase shift to converge in the spin wave, those generated from outside the stripes should have a phase delay instead of a phase advance compared with those generated from inside the stripes. This non-intuitive relationship indicates that the anisotropic dispersion of a spin wave is considered to be a dominating factor in setting the initial phase and position of a spin wave required to manipulate the waveform.

In conclusion, we have shown the unidirectional propagation of spin waves by generating them with circularly polarized light pulses focused on two stripe spots and allowing them to interfere. The propagation direction of the spin wave was switched by the polarization helicity of the pump pulses. The propagation asymmetry or unidirectionality of the spin wave was measured at various spacings of the two pump stripes. We confirmed that the spin waves propagate mostly unidirectionally when the stripe spacing is about $1/4$ of the wavelength. This phase manipulation technique was extended to a phased array of spin waves. Tuning the initial phase of the generated spin waves more finely (e.g., utilizing linearly polarized light pulses\cite{Yoshimine2014}) and designing phase difference, spatiotemporal design of the spin wave was shown to be achievable in simulations. This achievement opens up a field of magnetic materials science and explores an alternative sensing technique using magnetic fields.

The authors thank Y. Ozeki for the helpful consultation. This work was supported by grants from the Murata Science Foundation, Mitsubishi Foundation, KAKENHI (Grant Nos. 15H05454 and 26103004), and JSPS Core-to-Core Program, A. Advanced Research Networks.


\end{document}